\documentclass[
   ,final            
  ]
  {aipproc}

\usepackage{graphicx}
\layoutstyle{6x9}

\def\be{\begin{equation}}
\def\ee{\end{equation}}
\def\bea{\begin{eqnarray}}
\def\eea{\end{eqnarray}}

\def\ubar{{\bar u}}
\def\dbar{{\bar d}}
\def\sbar{{\bar s}}

\begin{document}

\title{Strangeness in the Meson Cloud Model}


\author{A. I. Signal}{
  address={Institute of Fundamental Sciences PN461 \\ Massey University \\ Palmerston North 4442 \\ New Zealand}
}

\begin{abstract}

I review progress in calculating strange quark and antiquark distributions of the nucleon using the meson cloud model.
This progress parallels that of the meson cloud model, which is now a useful theoretical basis for understanding symmetry breaking in nucleon parton distribution functions. 
I examine the breaking of symmetries involving strange quarks and antiquarks, including quark - antiquark symmetry in the sea, SU(3) flavour symmetry and SU(6) spin-flavour symmetry.

\end{abstract}

\maketitle


\section{Beginnings - Quark-Antiquark Asymmetry}

Tony Thomas and I met when I came to Adelaide as a new PhD student in early 1985. 
We agreed that I would work in the area of deep inelastic scattering, and in the first year Tony gave me a number of projects to work on. 
One of these was to try to extend his work from 1983 on the role of the non-perturbative pion cloud of the nucleon in DIS \cite{Tho83} to include kaons. 
We soon realized that the strangeness carrying components of the cloud would have different characteristics to the non-strange components.
This is because all the $\sbar$ antiquarks in the cloud come from the kaon, whereas all the $s$ quarks come from the hyperons. 
So immediately we saw the possibility that quark and antiquark could have different momentum distributions in the cloud. 
This was one of the first calculations to take into account the contributions to nucleon quark distribution functions coming from baryons in the cloud via the Sullivan process \cite{Sull72}, see fig. 1. 

\begin{figure}[b]
\centering
  \includegraphics[width=2.5in]{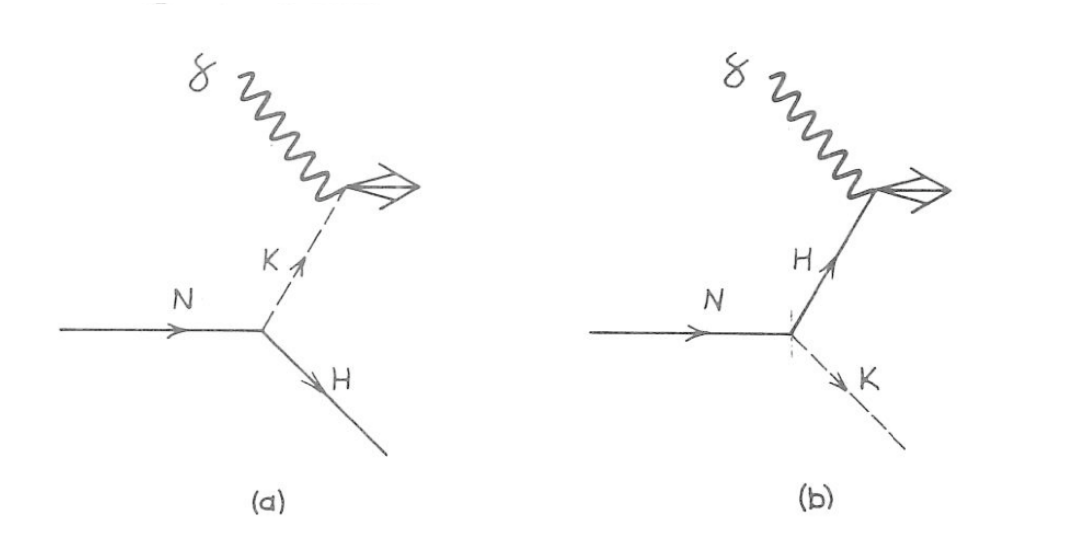}
  \hfill
  \caption{Non-perturbative contributions to the strange sea of the nucleon. 
  (a) The incoming photon is absorbed by a virtual kaon. 
  (b) The in coming photon is absorbed by a virtual hyperon}
\end{figure}

We were able to show that the contributions to the quark and antiquark distributions are given by convolutions between the distribution functions of quarks or antiquarks in the hyperon or kaon with the momentum distribution, or fluctuation function, of these hadrons in the cloud: 
\bea
x \delta \sbar(x) = \int_{x}^{1} dy\, f_{K}(y) \left(\frac{x}{y}\right) \sbar_{K}\left(\frac{x}{y}\right), \\
x \delta s(x) = \int_{x}^{1} dy\, f_{H}(y) \left(\frac{x}{y}\right) s_{H}\left(\frac{x}{y}\right).
\label{eq:sconv}
\eea

Using covariant perturbation theory we found \cite{ST87} that the meson cloud contribution to the antistrange distribution is softer than the contribution to the strange distribution. 
This arises mainly because we used a $\sbar_{K}$ distribution in the kaon that was fairly soft, and the fluctuation function
\bea
f_{K}(y) = f_{H}(1-y) 
\label{eq:cmf}
\eea
is also softer for the kaon than the hyperons.

As this was the first attempt to calculate strange contributions from the meson cloud, there were a number of shortcomings with this paper.
The first was the use of the covariant formulation  of perturbation theory in the calculation.
Unfortunately, using this formulation required us to make {\em ansatze} for the structure functions of the struck, off-shell, hadrons. 
We chose these to be the same as on-shell structure functions, which is not correct \cite{MST94}. 
A better formulation to use for the meson cloud model is time ordered perturbation theory in the infinite momentum frame, as shown by Wally Melnitchouk and Tony Thomas in an important paper for the development of the model \cite{MT93}.
A similar approach was also used by Zoller \cite{Zoll92}.
Using the time ordered approach has the advantages that the struck hadrons remain on-mass-shell, so avoiding any ambiguities and allowing us to use experimental input to the structure functions.
Also the momentum distributions in the cloud can be shown to satisfy the relation (\ref{eq:cmf}) exactly, rather than this being imposed by fiat.
In the infinite momentum frame, diagrams where the struck hadron is moving backwards in time are suppressed by powers of the longitudinal momentum, and do not contribute as the limit $p_{L} \rightarrow \infty$ is taken. 

We also had to make educated guesses for the strange and antistrange distributions in hyperons and kaons respectively. 
For the kaon we used an experimental determination of the pion structure function \cite{Bad83} which is fairly soft, whereas  for the hyperons we used a simple valence distribution of the form $s_{H}(x) = N_{s} x^{-1/2} (1-x)^{3}$. 
There was also no $Q^{2}$ dependence of our input or output distributions.

The question of  a possible quark - antiquark asymmetry in the strange sea received new interest in the early 2000's as a result of the interesting experimental result from the NuTeV collaboration \cite{NuTeV02}. 
NuTeV measured NC to CC ratios in deep-inelastic $\nu (\bar{\nu})$ - nucleon scattering. 
This enabled them to determine the effective couplings to left and right-handed quarks ($g_{L}$ and $g_{R}$) 
and, via the Paschos - Wolfenstein (PW) ratio, 
\bea
R_{PW} = \frac{\sigma^{\nu}_{NC} - \sigma^{\bar{\nu}}_{NC}}
{\sigma^{\nu}_{CC} - \sigma^{\bar{\nu}}_{CC}} = g_{L}^{2} - g_{R}^{2} = 
\frac{1}{2} - \sin^{2}\theta_{W},
\eea
the value of the weak mixing angle
\bea
\sin^{2}\theta_{W} = 0.2277 \pm 0.0013 (stat) \pm0.0009 (syst),
\eea
which is 2\% smaller than the world average value, or a $3\sigma$ discrepancy. 
However, the PW ratio receives corrections from both charge symmetry breaking in the nucleon parton distributions (which Tim Londergan and Tony Thomas have investigated in detail \cite{LT03}), and quark - antiquark symmetry breaking in the sea:
\bea
R_{PW} & = & \frac{1}{2} - \sin^{2}\theta_{W} + \nonumber \\
& & \frac{3b_{1}+b_{2}}{\langle x(u_{V}+d_{V})\rangle/2}  
\left[ - \langle x(s-\bar{s})\rangle + 
\frac{1}{2}(\langle x\delta u_{V}\rangle - \langle x\delta d_{V}) \rangle \right] 
\eea
where 
\begin{equation}
\delta u_{V} = u^{p}_{V} - d^{n}_{V};\;\;\delta d_{V} = d^{p}_{V} - u^{n}_{V}
\end{equation}
are the charge symmetry breaking valence distributions and 
\begin{equation}
b_{1} = \Delta_{u}^{2} = g_{L_{u}}^{2} - g_{R_{u}}^{2};\;\;
b_{2} = \Delta_{d}^{2} = g_{L_{d}}^{2} - g_{R_{d}}^{2}.
\end{equation}
At the NuTeV scale ($Q^{2} = 16$ GeV$^{2}$) the coefficient in front of the square 
brackets of eqn. (6) is  about 1.3, so a symmetry breaking term inside the square 
brackets of $-0.0038$ would explain the discrepancy between the NuTeV value and the 
accepted value of $\sin^{2}\theta_{W}$.
We note that the CTEQ group has analyzed the uncertainties around the experimental results for strange and anti-strange distributions in some detail \cite{CTEQ6.5S0}. 
They place bounds on the second moment of the quark - antiquark asymmetry
\bea
-0.001 < \langle x(s-\bar{s})\rangle <0.005.
\eea

This provided impetus to revisit our calculation of the asymmetry.
Now we do the calculation using time ordered perturbation theory, with on-shell structure functions. 
For the strange distribution in the hyperons we now use a bag model calculation \cite{BT+}, evolved using next-to-leading order QCD evolution to $Q^{2} = 16$ GeV$^{2}$.
The valence $\sbar(x)$ distribution in the kaon is taken from the parameterization of the Dortmund group \cite{GRS99}. 
We also note that the form factors cutting off the $NKH$ vertex are fairly soft ($\Lambda_{c} \sim 1$ GeV).
One further point of difference to our original calculation is the inclusion of $K^{*}$ meson Fock states.
This can have a significant effect on the calculations, as the coupling constants for $K^{*}NH$ are fairly large \cite{HHoltmannSS}.
Also the fluctuation functions for $N \rightarrow K^{*}H$ peak close to $y = 0.5$, meaning that the final convolutions to obtain the contributions to $s$ and $\sbar$ reflect the underlying hardness or softness of the valence quark distribution in the hadron.
However, we realize that we are pushing the bounds of the cloud model here, as it is not clear that $K^{*}H$ final states would have a clear rapidity gap.

We find that the fluctuation functions for kaons are softer than for hyperons, whereas the $s$ quark distributions in $\Lambda$ and $\Sigma$ hyperons are now softer than the $\sbar$ distribution in the $K$ and $K^{*}$.
This means that once the quark distributions are convoluted with the fluctuation functions, there is only a small $s - \sbar$ difference, see fig. 2.
The second moment of the asymmetry has a magnitude around $10^{-4}$, and positive (negative) sign without (with) $K^{*}$ states included.
As this is significantly smaller than the size of effect needed to move the NuTeV result into agreement with the world data, we conclude that the strange sea asymmetry is probably not responsible for the NuTeV anomaly.

\begin{figure}[bt]
\centering
  \includegraphics[width=2.5in]{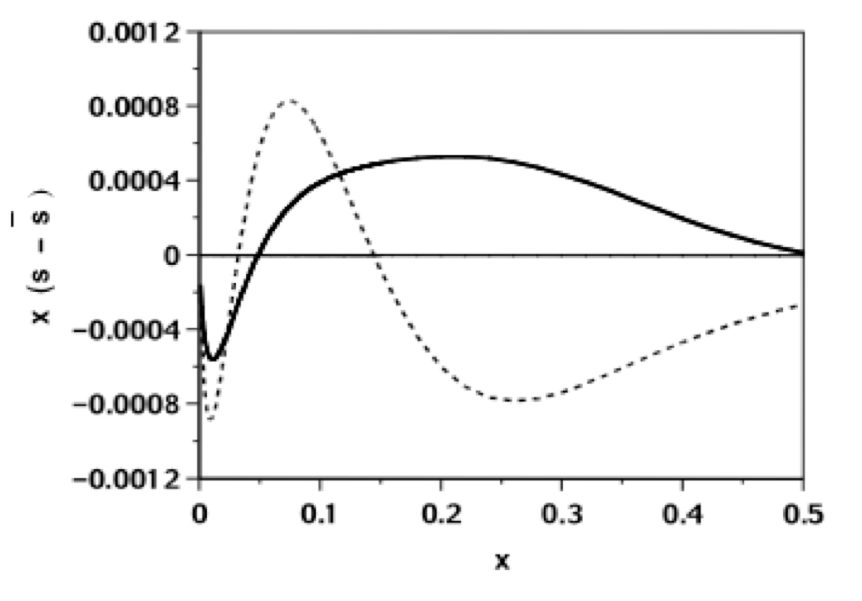}
  \hfill
  \caption{The strange sea asymmetry calculated in the meson cloud model. 
  The solid and dashed curves are the results without and with the $K^{*}$ contributions respectively.  }
\end{figure}

\section{Polarised Strange Sea}

The calculational techniques outlined in the above section can be generalized to the polarized quark distributions $\Delta s(x)$ and $\Delta \sbar(x)$ without too much difficulty.
Polarized quark distributions have been of interest for over 20 years, since the EMC collaboration measured a very small fraction of the nucleon spin being carried by quarks \cite{EMC88}.
This is usually interpreted in the context of SU(3) flavour and implies that the strange sea is strongly polarized opposite to the proton $\Delta S \simeq -0.15$ \cite{BEK88}.
It has been pointed out that a natural consequence of the meson cloud model is that the cloud is capable of carrying a significant proportion of the proton's angular momentum \cite{ST97}. 

The HERMES collaboration have carried out an extensive programme of flavour analysis of their polarized DIS data \cite{Hermes04,Hermes08}, which shows that the polarized sea quark distributions are fairly small.
Our calculations in the MCM, which include the contributions from $K^{*}$ states, are consistent with HERMES data, see fig. 3.

\begin{figure}[bt]
\centering
  \includegraphics[width=2.5in]{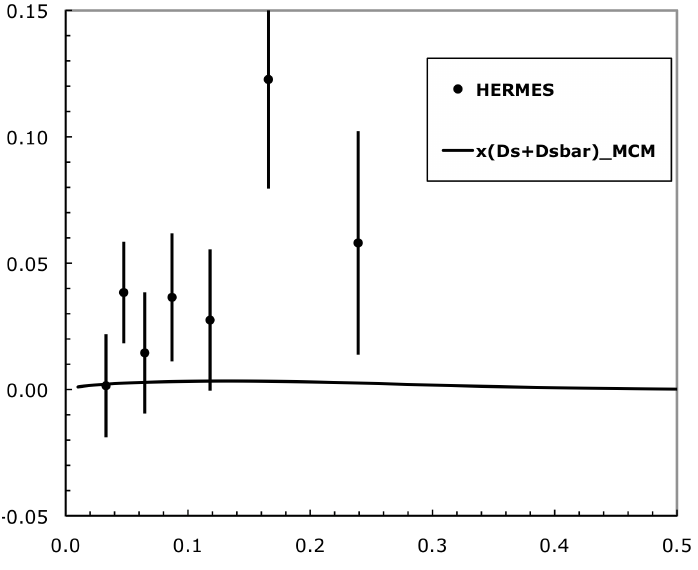}
  \hfill
  \caption{Comparison of MCM calculations for $x(\Delta s + \Delta \sbar)$ with the HERMES data at $Q^{2} = 2.5$ GeV$^{2}$.  }
\end{figure}

\section{SU(3) Flavour Symmetry Breaking}

The unpolarized strange sea is less well constrained by experimental data than the light ($\ubar, \dbar$) sea.
For instance the CTEQ6.5 pdf set \cite{CTEQ6.5S0} has a very large variance in the parameters describing the $s$ and $\sbar$ distributions ($\pm$ 50\% in some instances). 
The recent HERMES data \cite{Hermes08} on the strange sea highlights this problem, as it does not agree well with the NuTeV determination \cite{NuTeV07} - though we note that the HERMES analysis of their data is only to leading order in QCD, whereas the NuTeV analysis goes to next-to-leading order.

In the MCM, we can estimate the strange sea via the SU(3) flavour breaking asymmetry 
\bea
\Delta(x) = \ubar(x) + \dbar(x) - s(x) - \sbar(x)
\eea
which has leading contributions in the cloud coming from the differences between {\em e.~g.~} 
$|N \pi \rangle - |\Lambda K \rangle$ Fock states.
On the other hand, there are no leading contributions to $\Delta(x)$ in perturbative QCD (and next-to-leading contributions can also be expected to be small).
Having calculated $\Delta(x)$ in the MCM, we can subtract the light sea distributions, which are experimentally well constrained, and estimate the total strange sea. 
Our results are shown in fig. 4, and are generally consistent with the HERMES data. 
We have again included $K^{*}$ states in the calculation of $\Delta(x)$, but they do not dominate the final results, and removing them has about a 10\% effect on our total $s(x) + \sbar(x)$. 
We note that our calculation becomes negative at $x \approx 0.25$, which is unphysical.
This could be due to either the MCM calculation overestimating $\Delta(x)$ or the CTEQ6.6 pdf set \cite{CTEQ6.6} underestimating the light sea $[ \ubar(x) + \dbar(x) ]$  or both.

In conclusion, the meson cloud model remains an excellent non-perturbative laboratory for exploring and understanding symmetry breaking among the nucleon parton distribution functions.
There are still important questions around the polarized and unpolarized strange sea distributions, andthe model can help to provide solutions to these.

\begin{figure}[bt]
\centering
  \includegraphics[width=2.5in]{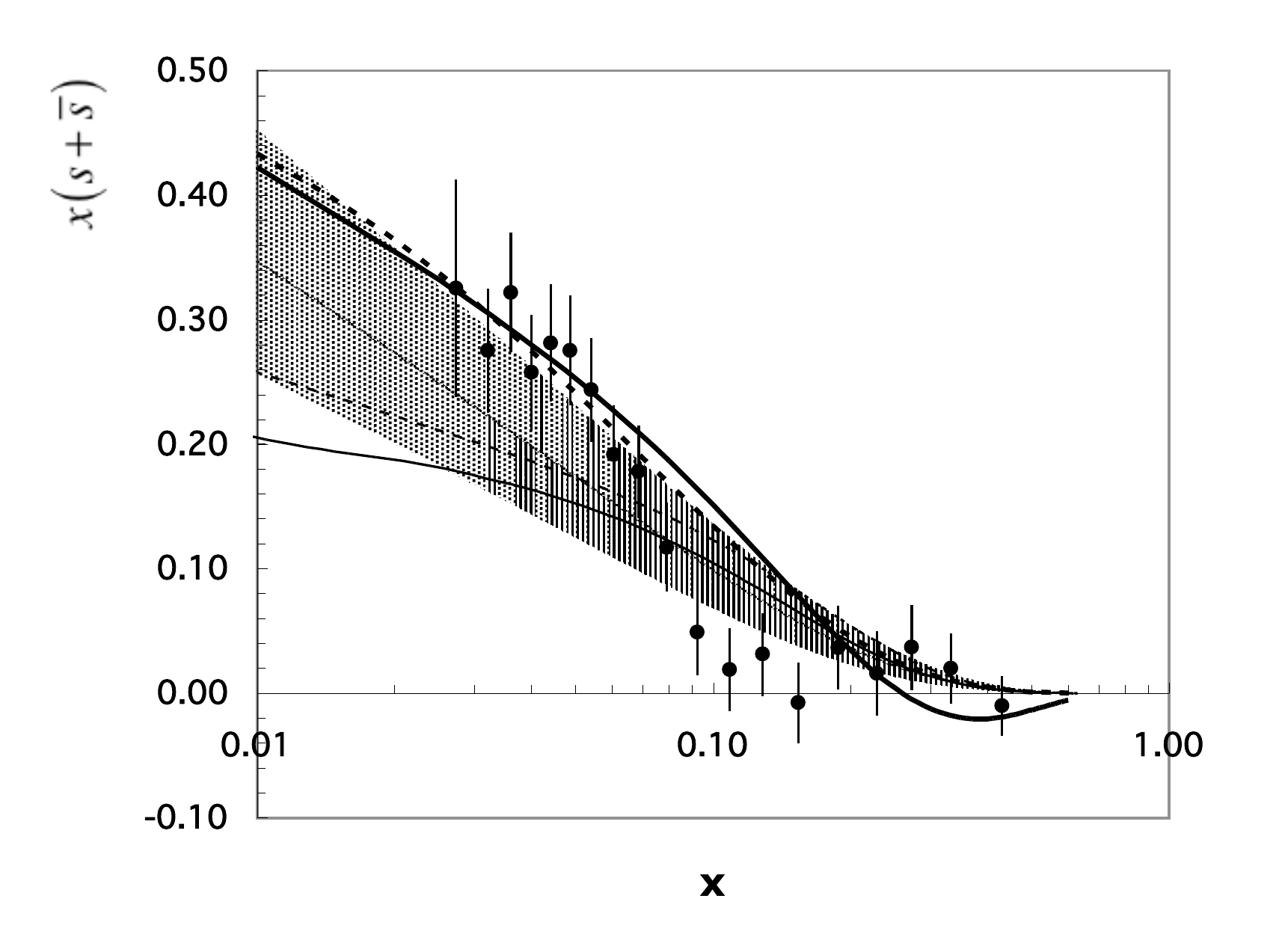}
  \hfill
  \caption{The sum of the strange and antistrange quark distributions from the MCM calculations (the thick solid curve), the HERMES measurements (the data points) and  the global fit results from CTEQ6.6M (the thick dashed curve), MSTW2008 (the dash curve) and CTEQ6.5 (the shaded area), and the next-to-leading order analysis of NuTeV dimuon data 
(the thin solid curve).  }
\end{figure}


\begin{theacknowledgments}
I am happy to acknowledge the contributions to my understanding of the meson cloud model that have come from many colleagues and friends.
Firstly to Tony Thomas, who introduced me to this problem, guided me through my PhD studies, and has been incredibly generous with his time and support over 25 years.
Also I have learnt a great deal from other Adelaide students especially Andreas Schreiber, Wally Melnitchouk and Fernanda Steffans.
My colleagues at Massey University, Fu-Guang Cao and Francois Bissey, have provided many insights, and I am grateful to them for many years of enjoyable collaboration.
\end{theacknowledgments}



\bibliographystyle{aipproc}   


\begin{thebibliography}{9}

\bibitem{Tho83}
	A. W. Thomas, \emph{Phys. Lett. B}, \textbf{126}, 97 (1983);
	M. Ericson and A.W. Thomas, \emph{Phys. Lett. B}, \textbf{128}, 122 (1983).

\bibitem{Sull72}
	J. D. Sullivan, \emph{Phys. Rev. D}, \textbf{5}, 1732 (1972).

\bibitem{ST87}
	A. I. Signal and A. W. Thomas, \emph{Phys. Lett. B}, \textbf{191}, 205 (1987).

\bibitem{MST94}
	W. Melnitchouk, A. W. Schreiber and A. W. Thomas, \emph{Phys. Rev. D}, \textbf{49}, 1183 (1994).

\bibitem{MT93}
	W. Melnitchouk and A. W. Thomas, \emph{Phys. Rev. D}, \textbf{47}, 3794 (1993).
	
\bibitem{Zoll92}
	V. Zoller, \emph{Z. Phys. C}, \textbf{53}, 443 (1992).

\bibitem{Bad83}
	J. Badier et al., \emph{Z. Phys. C}, \textbf{18}, 291(1983).

\bibitem{NuTeV02} 
	G. P. Zeller et al. (NuTeV collaboration),  \emph{Phys. Rev. Lett.}, \textbf{88}, 091802 (2002).
	
\bibitem{LT03}
	J. T. Londergan and A. W. Thomas, \emph{Phys. Rev. D}, \textbf{67}, 111901(R) (2003).
	
\bibitem{CTEQ6.5S0}
	H. L. Lai, et. al. (CTEQ collaboration), \emph{J. High Energy Phys.}, \textbf{0704}, 089 (2007).

\bibitem{BT+}
	C. Boros and A. W. Thomas,  \emph{Phys. Rev. D}, \textbf{60}, 074017 (1999); 
	F. G. Cao and A. I. Signal, \emph{Phys. Lett. B}, \textbf{474}, 138 (2000); 
	F. G. Cao and A. I. Signal, \emph{Phys. Lett. B}, \textbf{559}, 229 (2003). 
	
\bibitem{GRS99}
	M. Gl\"{u}ck, E. Reya and I. Schienbein, \emph{Eur. Phys. J. C}, \textbf{10}, 313 (1999).

\bibitem{HHoltmannSS}
	H. Holtmann, A. Szczurek and J. Speth, \emph{Nucl. Phys. A}, \textbf{569}, 631 (1996).
	
\bibitem{EMC88}
	J. Ashman et al. (EMC collaboration), \emph{Phys. Lett. B}, \textbf{206}, 364 (1988).
		
\bibitem{BEK88}
	S. J. Brodsky, J. Ellis and M. Karliner, \emph{Phys. Lett. B}, \textbf{206}, 309 (1988);
	S. D. Bass, \emph{The Spin Structure of the Proton}, World Scientific, Singapore, 2007.
	
\bibitem{ST97}
	J. Speth and A. W. Thomas, \emph{Adv. Nucl. Phys.}, \textbf{24}, 83 (1997); 
	F. Bissey, F. G. Cao and A. I. Signal,  \emph{Phys. Rev. D}, \textbf{73}, 094008 (2006).

\bibitem{Hermes04}
	A. Airapetian et al. (HERMES collaboration), \emph{Phys. Rev. Lett.}, \textbf{92}, 012005 (2004).
	
\bibitem{Hermes08}
	A. Airapetian et al. (HERMES collaboration), \emph{Phys. Lett. B}, \textbf{666}, 446 (2008).

\bibitem{NuTeV07} 
	D. Mason et al. (NuTeV collaboration),  \emph{Phys. Rev. Lett.}, \textbf{99}, 192001 (2007).

\bibitem{CTEQ6.6}
	P. M. Nadolsky, et. al. (CTEQ collaboration), \emph{Phys. Rev. D}, \textbf{78}, 013004 (2008).

\end{thebibliography}


\end{document}